\documentclass[fleqn,usenatbib]{mnras}

\usepackage{newtxtext,newtxmath}
\usepackage[T1]{fontenc}

\DeclareRobustCommand{\VAN}[3]{#2}
\let\VANthebibliography\thebibliography
\def\thebibliography{\DeclareRobustCommand{\VAN}[3]{##3}\VANthebibliography}

\usepackage{graphicx}   % Including figure files
\usepackage{amsmath}    % Advanced maths commands
\usepackage{bm}

\usepackage{amssymb}    % Extra maths symbols

\newcommand{\kms}{{\rm \,km/s}}

\title[Constraints on the velocity dispersion function]
{The velocity dispersion function of early-type galaxies and its
redshift evolution: the newest results from lens redshift test}

\author[Geng et al.]{
Shuaibo Geng,$^{1}$ Shuo Cao,$^{1}$\thanks{caoshuo@bnu.edu.cn}
Yuting Liu,$^{1}$ Tonghua Liu,$^{1}$ Marek Biesiada,$^{1,2}$ Yujie
Lian$^{1}$
\\
$^1$ Department of Astronomy, Beijing Normal University, 100875,
Beijing, China; \\
$^2$ National Centre for Nuclear Research,
Pasteura 7, 02-093 Warsaw, Poland
}

\date{Accepted XXX. Received YYY; in original form ZZZ}

\pubyear{2020}

\begin{document}
\label{firstpage}
\pagerange{\pageref{firstpage}--\pageref{lastpage}}
\maketitle

\begin{abstract}
The redshift distribution of galactic-scale lensing systems provides
a laboratory to probe the velocity dispersion function (VDF) of
early-type galaxies (ETGs) and measure the evolution of early-type
galaxies at redshift $z\sim 1$. Through the statistical analysis of
the currently largest sample of early-type galaxy gravitational
lenses, we conclude that the VDF inferred solely from strong lensing
systems is well consistent with the measurements of SDSS DR5 data in
the local universe. In particular, our results strongly indicate a
decline in the number density of lenses by a factor of two and a
20\% increase in the characteristic velocity dispersion for the
early-type galaxy population at $z\sim1$. Such VDF evolution is in
perfect agreement with the $\Lambda$CDM paradigm (i.e., the
hierarchical build-up of mass structures over cosmic time) and
different from "stellar mass-downsizing" evolutions obtained by many
galaxy surveys. Meanwhile, we also quantitatively discuss the
evolution of the VDF shape in a more complex evolution model, which
reveals its strong correlation with that of the number density and
velocity dispersion of early-type galaxies. Finally, we evaluate if
future missions such as LSST can be sensitive enough to place the
most stringent constraints on the redshift evolution of early-type
galaxies, based on the redshift distribution of available
gravitational lenses.
\end{abstract}


%%%%%%%%%%%%%%%%%%%%%%%%%%%%%%%%%%%%%%%%%%%%%%%%%%

%%%%%%%%%%%%%%%%% BODY OF PAPER %%%%%%%%%%%%%%%%%%

\section{Introduction}

The formation and evolution of early-type galaxies (ETGs) have been
the focus of many observational studies, because they can provide
more robust tests of the underlying $\Lambda$CDM theory. Meanwhile,
one of the most interesting problems of the early-type galaxies is
to determine their velocity dispersion function (VDF), which can
provide clues to galaxy formation and evolution. In the last decade,
the wealth of data from large sky surveys such as Sloan Digital Sky
Survey (SDSS; \citep{Abazajian09}) enabled the determination of the
fundamental VDF parameters at different scales, based on the
spectroscopic measurements of stellar kinematics within the
effective radius, as well as the extended X-ray emitting gas
temperature extended to the dark halo. Specifically, in the local
universe the VDF has been measured using direct kinematic
measurements from the SDSS spectroscopic data of early-type galaxies
\citep{Sheth03,Choi2007}. However, for a given large sample of
galaxies such as the SDSS sample, accurately classifying large
numbers of galaxies is the major difficulty in deriving reliable
type-specific VDF \citep{Chae2007}. The low-velocity dispersion bias
should also be appropriately corrected, i.e., in the magnitude
limited catalogs the galaxy number counts become incomplete at
low-velocity dispersions \citep{Bernardi2010}. On the other hand,
the shape and redshift evolution of the VDF carries information
about the physical mechanisms responsible for the growth of a
galaxy. For instance, the central gas accretion and the resulting
star formation could efficiently increase the velocity dispersion
and mass, while mass loss in galactic winds could play a different
role. Whereas, the evolutions of the VDF from a high redshift
Universe are not understood as well, although early-type galaxies,
especially massive elliptical galaxies, are commonly thought to be
the end results of the galaxy merging and accretion processes
\citep{Kauffmann1993,Cole1994,Kauffmann1996} (however see
\citep{Renzini06} for a far more complicated history). Observational
evidence of such scenarios can be directly compared to predictions
from cosmological simulations, based on the SDSS luminosity
functions and intrinsic correlations between luminosity and velocity
dispersion \citep{Chae2010}.

Independent of the traditional redshift surveys, in this work we
constrain the VDF of early-type galaxies using the statistics of
strong gravitational lensing systems
\citep{Turner1984,Biesiada06,Cao12b,Cao12c}. Assuming the
concordance cosmological model ($\Lambda$CDM), several efforts have
been made to include the distribution of lensed image separations in
the study of the galaxy mass profiles and the evolution history of
galaxies. Based on the CLASS and PANELS lens sample, the first
attempt to constrain the redshift evolution of galaxies (since
redshift $z\sim 1$) was presented in \citet{Chae2003}. This study
was then extended to the study of the shape of the VDF and the
characteristic velocity dispersion \citep{Chae05}. However, the
sample size of the data available at that time did not allowed for a
firm determination of the galaxy velocity dispersion function. Here
we present a new approach to derive the VDF based on the lens
redshift distribution \citep{Kochanek1992} and to constrain its
evolution out to $z\sim 1$, given its strong dependency on the
dynamical properties of galaxies (i.e., stellar velocity dispersion)
and the number density of gravitational lenses (i.e., galaxy
evolution). Compared with the previous works focusing on image
separation distributions, constraining a VDF through lens redshift
test is unique and promising, since it does not require the
knowledge of the total lensing probability and the magnification
bias in the sample \citep{Ofek2003}. The advantages of the lens
redshift test have been extensively discussed in
\citep{Matsumoto2008,Koopmans2009,Oguri2012,Cao12a}. Therefore, it
will be rewarding to investigate the velocity dispersion function
and evolution of the lensing galaxies by adopting the cosmological
parameters determined by \textit{Planck} and using a new lens sample
better representing the distribution of the galaxy properties. In
this work, we focus on a newly compiled sample of 157 galaxy-scale
strong lensing systems, which are all early-type lenses (E or S0
morphologies) without significant substructures or close companion
galaxies \citep{Chen19}. Throughout the paper we assume the
concordance cosmology by adopting the cosmological parameters
determined by the \textit{Planck} 2016 data \citep{Ade2016}.

\section{Methodology and observational data}

Because early-type galaxies dominate the lensing cross sections due
to their large central mass concentrations \citep{Keeton1997}, one
may naturally expect them being a unique mass-selected sample to
study the VDF and evolution of galaxies, which has triggered
numerous efforts to use early-type galactic lenses for this purpose.
We mostly follow the methodology described in
\citet{Ofek2003,Cao12a} to calculate the lensing probabilities for
the lens sample, although a number of modifications and updates are
included for more accurate calculations. Based on the number density
of the lens $n(\theta_E,z_l)$ and the lensing cross-section $S_{cr}$
for multiple imaging, the differential probability that a source
with redshift $z_s$ will be multiply imaged with Einstein radius
$\theta_E$ by a distribution of galaxies per unit redshift can be
defined by
\begin{equation}
\frac{d\tau}{dz_l}=n(\theta_E,z_l)(1+z_l)^{3} S_\mathrm{cr}
\frac{cdt}{dz_l} \, , \label{opt_depth}
\end{equation}
Here the lensing cross-section and the proper distance interval are
respectively expressed as
\begin{equation}
S_\mathrm{cr}=\pi \theta_E^2 D_{l}^2,
\end{equation}
and
\begin{equation}
\frac{cdt}{dz_l}=\frac{c}{(1+z_l)}\frac{1}{H(z_l)},
\end{equation}
with $H(z)$ and $D_{l}$ representing the expansion rate of the
Universe at redshift $z$ and the angular diameter distance between
the observer and the lens. In this paper, following the method
proposed by \citep{Ofek2003}, we use the differential optical depth
for lensing with respect to the lens redshift $z_l$ as the
probability density. Then the relative probability of finding the
early-type lens at redshift $z_l$ for a given source with Einstein
radius $\theta_E$ is derived as
\begin{equation}
\delta p=\frac{d\tau}{dz_l}/{\tau}=\frac{d\tau}{dz_l}/\int_0^{z_s}
\frac{d\tau}{dz_l} dz_l.
\end{equation}

I. The velocity dispersion function of galaxies (or the lens number
density) is an essential part of the theoretical prediction of the
lensing probability. Assuming a power-law relation between
luminosity ($L$) and velocity dispersion ($\sigma$), the
distribution of early-type galaxies in velocity dispersion can be
described by the modified Schechter function
\begin{equation}
\frac{dn}{d\sigma}=n_{\ast} (\frac{\sigma}{\sigma_{\ast}})^{\alpha}
{\rm exp}\left[-(\frac{\sigma}{\sigma_{\ast}})^{\beta}\right]
\frac{\beta}{\Gamma(\alpha/\beta)} \frac{1}{\sigma}
\end{equation}
where $n_{\ast}$ is the integrated number density of galaxies, and
$\sigma_{\ast}$ is the characteristic velocity dispersion. The shape
of VDF is characterized by the low-velocity power-law index
($\alpha$) and the high-velocity exponential cut-off index
($\beta$). Note that $\alpha$ and $\beta$ are not only used as
important input to the strong lensing statistical analysis, but also
contribute to revealing the features of the distribution  of
early-type galaxies. Thus, in this paper an independent method --
the redshift information of the lensing galaxies -- will be used to
place constraints on the VDF shape parameters. We also consider the
possibility of redshift evolution of the velocity function through a
parametric approach, which is supported by the high-resolution
N-body simulation following the evolution of ${512}^3$ particles in a
cosmological box of ${100h}^{-1}$ Mpc \citep{Jing02}. More
specifically, the evolutions of the number density and the velocity
dispersion are parameterized as
\begin{equation}
\begin{aligned}
n_{\ast}(z_l)\rightarrow n_{\ast}(1+z_l)^{\nu_n},
\sigma_{\ast}(z_l)\rightarrow \sigma_{\ast}(1+z_l)^{\nu_v},
\end{aligned}
\end{equation}
which is a power-law evolution model extensively discussed
in the previous analysis of lens statistics
\citep{Chae2003,Matsumoto2008, Oguri2012}. The no-evolution model
is quantified by ($\nu_{n}=\nu_{v}=0$) and has been adopted in most
of previous studies, while the cases of ($\nu_{n}<0$, $\nu_{v}>0$)
and ($\nu_{n}>0$, $\nu_{v}<0$) correspond to two different views of
number and mass evolution in the early-type population from $z=0$ to
1. In this paper, we also focus on another typical redshift
evolution model proposed by \citet{Ofek2003,Chae2010}, which allows
the number density of galaxies and the velocity dispersion to vary
with redshift as
\begin{equation}
\begin{aligned}
n_{\ast}(z_l)\rightarrow n_{\ast} 10^{Pz_l}, \sigma_{\ast}(z_l)
\rightarrow \sigma_{\ast}10^{Qz_l}.
\end{aligned}
\end{equation}
Here $P$ and $U$ are two constant quantities to be determined from
the redshift distribution of galactic-scale lensing systems (see the
"Discussion" section for details).

II. The lens potential is assumed to originate from a spherically
symmetric power-law mass distribution $\rho\sim r^{- \gamma}$
\citep{Treu2006,Cao2015,Pan2016,Qi2019}, in the framework of which
the characteristic Einstein radius can be expressed as
\begin{equation}
\theta_{E\ast}=\lambda(e) [4\pi (\frac{\sigma_{ap}}{c})^2
\frac{D_{ls}}{D_s}{\theta_{ap}}^{\gamma-2}
f(\gamma)]^{\frac{1}{\gamma-1}}.
\end{equation}
Here $D_{ls}$ and $D_{s}$ are the angular diameter distances between
lens-source and observer-source, respectively. Note that
$\lambda(e)$ denotes a dynamical normalization factor
\citep{Keeton1998}, while $f(\gamma)$ is a function of the radial
mass profile slope and $\sigma_{ap}$ is the luminosity averaged
line-of-sight velocity dispersion inside the aperture $\theta_{ap}$
\citep{Cao2015}. The power-law model can be derived by solving the
spherical Jeans equation analytically assuming that stellar and
total mass distributions follow the same power-law and velocity
anisotropy vanishes \citep{Koopmans05}, which has been widely used
in several studies of lensing events caused by early-type galaxies
\citep{Treu2002,Treu2006,Cao2016,Cao20,Cao21}. In this analysis, we
adopt the latest constraints on the average logarithmic density
slope, based on the direct total-mass and stellar-velocity
dispersion measurements from a large sample of secure strong
gravitational lens systems \citep{Koopmans2009}. Considering the
three dimensional shapes of lensing galaxies one can take the
normalization factor as a mean of two equally probable extreme cases
(oblate and prolate):
$\lambda(e)=0.5\lambda_{obl}(e)+0.5\lambda_{pro}(e)$, where the
partial normalization factors are parameterized in the form of
\citep{Oguri2012}, with the ellipticity derived from the axis ratio
distributions of early-type galaxies in the SDSS survey
\citep{Bernardi2010}.

In this paper, following the method proposed by \citet{Ofek2003}, we
use the differential optical depth to lensing with respect to the
lens redshift $z_l$ as the probability density. For a statistical
sample that contains $N_l$ strong lensing systems, the
log-likelihood of observing the lens at redshift $z_l$ is given by
\begin{equation}
{\rm ln}\mathcal{L}(\textbf{p})=\sum_{i=1}^{N_l} {\rm ln} \delta
p_i(\textbf{p}),
\end{equation}
where \textbf{p} represents the set of the velocity dispersion
function parameters ($\alpha$, $\beta$) and the galaxy evolution
parameters ($\nu_n$, $\nu_v$). Now one can perform Monte Carlo
simulations of the posterior likelihood ${\cal L} \sim \exp{(-
\chi^2 / 2)}$, where the $\chi^2$ function is defined as
\begin{equation}
\chi^2=-2{\rm ln}\mathcal{L}.
\end{equation}
in our statistical analysis of lens redshift distribution. The
sample used in this paper is primarily drawn from Sloan Lens ACS
Survey (SLACS) and recent large-scale observations of galaxies,
which is compiled and summarized in \citet{Cao2015,Shu2017}. The
combined sample includes 91 lenses from SLACS \citep{Shu2017,
Bolton2008,Auger2009} and an extension of the SLACS survey known as
``SLACS for the Masses" (S4TM) \citep{Shu2015,Shu2017}, 35 lenses
from the BOSS emission-line lens survey (BELLS)
\citep{Brownstein2012} and BELLS for GALaxy-Ly$\alpha$ EmitteR
sYstemsGALLERY (BELLS GALLERY) \citep{Shu16a,Shu16b}), 26 lenses
from the Strong Lensing Legacy Survey (SL2S)
\citep{Sonnenfeld13a,Sonnenfeld13b}, and 5 lenses from Lenses
Structure and Dynamic (LSD) \citep{Treu2002,Koopmans2003,Treu2004}.
The advantage of this recently assembled lens sample, the detailed
information of which is described and is listed in \citet{Chen19},
lies in its well-defined observational selection criteria satisfying
the assumption of spherical lens mass model. Fig.~1 shows the
redshift distributions of the lensing systems used in our analysis.
However, a statistical analysis requires a sample that is complete
and has well characterized, homogeneous selection criteria. Note
that the lensing systems collected in this analysis are selected in
very different manners. For instance, the SLACS, S4TM, and BELLS
surveys respectively selected candidates from the spectroscopic
observations of early-type galaxies and look for the presence of
higher-redshift emission lines in Sloan Digital Sky Survey I
\citep{Eisenstein01} and Sloan Digital Sky Survey-III
\citep{Eisenstein11}. These lens candidates were followed up with
HST ACS snapshot imaging and after image processing. Therefore, in
order to verify the completeness of the full early-type lens sample
(hereafter Sample A), one additional sub-sample will also be applied
to discuss its utility for the redshift test: 126 deflector-selected
lenses from SLACS, S4TM, BELLS and BELLS GALLERY (hereafter Sample
B). Such choice is also motivated by the fact that the SLACS and
BELLS lenses could be moderately suffered from the finite Sloan
fiber size \citep{Brownstein2012}.

\begin{figure}
\centering
\includegraphics[width=10cm]{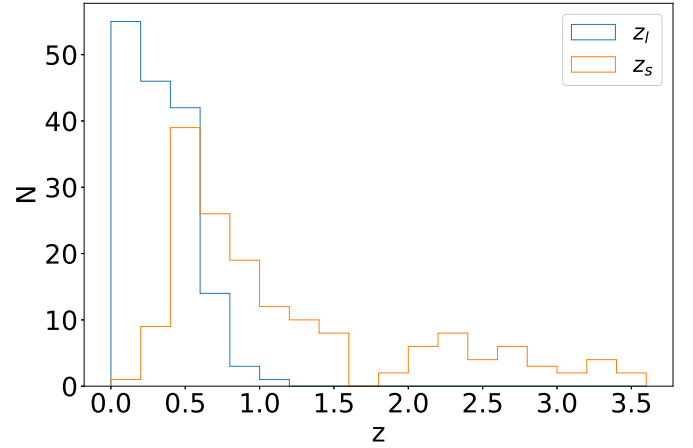}
\caption{The scatter plot of 158 strong lensing systems used in the
lens redshift test.} \label{fig:redshift}
\end{figure}

\section{Results}

\subsection{Constraints on the shape of velocity dispersion function}

We constrain the VDF of the form given by Eq.~(5) for the early-type
VDF, which is well-fitted by three effective parameters, i.e., the
characteristic velocity dispersion ($\sigma_*$), the low-velocity
power-law index ($\alpha$), and the high-velocity cut-off index
($\beta$) \footnote{The characteristic number density ($n_*$) has no
relation to the shape of lens redshift distribution, since a
relative lensing probability as a function of $z_l$ (see Eq.~(4))is
used in this analysis, instead of the absolute lensing probability
through Eq.~(1). This allows us to fix $n_*$ by an arbitrary
constant.}. However, the strong lensing systems number statistics is
not a strong enough test to enable the four-parameter function
\citep{Chae05}. Hence in order to break the parameter degeneracy
without significantly altering the possible range of the VDF, we
will focus on the constraints on shape of the VDF, with $\sigma_*$
fixed at the best-fit value by the SDSS DR5 local central stellar
VDF \citep{Choi2007}
\begin{eqnarray}
 ( \sigma_*,\hspace{0.2cm} \alpha,
 \hspace{0.2cm} \beta)_{\rm DR5}
  & =  & [161 \pm 5 \hspace{0.2cm} km/s, \nonumber \\
  &   & 2.32 \pm 0.10, \hspace{0.2cm} 2.67 \pm 0.07].
\label{DR5}
\end{eqnarray}
The numerical results for the VDF shape parameters are summarized in
Table 1, with the marginalized confidence limits (C.L.) on the
parameter plane ($\alpha, \beta$) presented in Fig.~1. Recent
measurements of three stellar VDFs are also added for comparison:
the VDF for local early-type galaxies based on SDSS Data Release 5
\citep{Choi2007}, an inferred local stellar VDF obtained through
Monte Carlo simulations, based on the galaxy luminosity functions
from the SDSS and intrinsic correlations between luminosity and
velocity dispersion \citep{Chae2010}, and the VDF for quiescent
galaxies in the local universe, based on the Main Galaxy Sample from
the SDSS Data Release 12 \citep{Sohn2017}.

\begin{figure*}
\centering
\includegraphics[width=12cm]{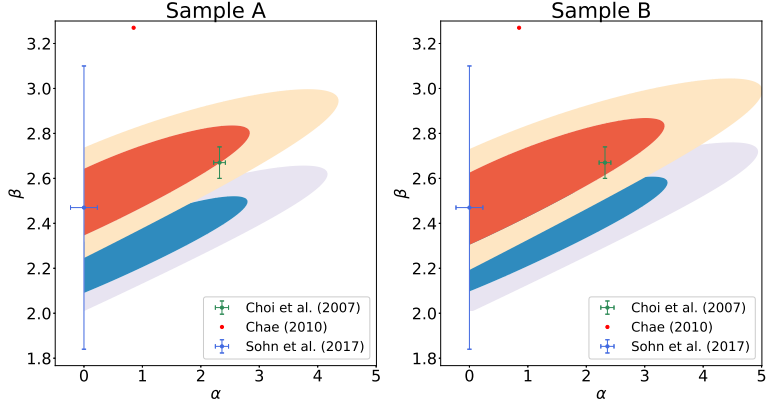}
\caption{Constraints on the shape of the VDF, i.e., low-velocity
power-law index $\alpha$ and high-velocity exponential cut-off index
$\beta$ with two lens samples (Sample A and Sample B), in the
framework of no-evolution (blue contours) and redshift-evolution
model (red contours). }
\end{figure*}

\begin{table*}
\caption{Summary of the constraints on the shape of the model VDF of
early-type galaxies, based on the lens redshift distribution of the
current strong lensing observations. }
    \label{tab:1}
    \begin{tabular}{l|c|c|c|c}
        \hline \hline
        VDF Evolution & Data & $\sigma_*$ (km/s) & $\alpha$ & $\beta$\\
        \hline
        $\nu_{n}=\nu_{v}=0$ & Sample A & 161 & $0.66^{+2.13}_{-0.66}$ & $2.28^{+0.24}_{-0.18}$ \\
        $\nu_{n}=\nu_{v}=0$ & Sample B & 161  & $1.00^{+2.38}_{-1.00}$ & $2.34^{+0.26}_{-0.24}$ \\
        $\nu_{n}=-1.0, \nu_{v}=0.25$ & Sample A & 161 & $0.45^{+2.38}_{-0.45}$ & $2.55^{+0.28}_{-0.20}$ \\
        $\nu_{n}=-1.0, \nu_{v}=0.25$ & Sample B & 161  & $0.70^{+2.63}_{-0.70}$ & $2.55^{+0.30}_{-0.24}$ \\
        \hline
        $\nu_{n}=\nu_{v}=0$ & Sample A & $161\pm5$ & $0.86^{+2.18}_{-0.86}$ & $2.30^{+0.24}_{-0.20}$ \\
        $\nu_{n}=-1.0, \nu_{v}=0.25$ & Sample A & $161\pm5$ & $0.48^{+2.40}_{-0.48}$ & $2.57^{+0.28}_{-0.21}$ \\
        \hline \hline
    \end{tabular}
\end{table*}

In the first scenario, we assume that neither the characteristic
velocity dispersion ($\sigma_*$) nor the number density ($n_*$) of
galaxies evolves with redshifts ($\nu_n=\nu_v=0$). Given the
redshift coverage of the lensing galaxies in the lens sample
($0.06<z_l<1.0$), if we constrain a non-evolving VDF using the lens
data, then, assuming the VDF evolution with redshift is smooth, the
fits on the VDF parameters may represent the properties of
early-type galaxies at an effective epoch of $z\sim 0.5$.  Such
non-evolving velocity dispersion function has been extensively
applied in the previous studies on lensing statistics
\citep{Chae2003,Ofek2003,Capelo2007,Cao12a}. By applying the above
mentioned $\chi^2$ - minimization procedure to Sample A, we obtain
the best-fit values and corresponding $1\sigma$ uncertainties
(68.3\% confidence level): $\alpha=0.66^{+2.13}_{-0.66}$,
$\beta=2.28^{+0.24}_{-0.18}$. It is obvious that the full sample
analysis has yielded improved constraints on the high-velocity
exponential cut-off index $\beta$, compared with the previous
analysis of using the distribution of image separations observed in
CLASS and PANELS to constrain a model VDF of early-type galaxies
\citep{Chae05}. Suffering from the limited size of lens
sample, such analysis \citep{Chae05} found that neither of the two
VDF parameters ($\alpha$, $\beta$) can be tightly constrained, due
to the broad regions in the $\alpha-\beta$ plane. Consequently, the
image separation distribution is consistent with the SDSS measured
stellar VDF \citep{Sheth03} and the Second Southern Sky Redshift
Survey (SSRS2) inferred stellar VDF \citep{Chae2003}, although the
two stellar VDFs are significantly different from each other
concerning their corresponding parameter values. We also consider
constraints obtained for the Sample B (defined in previous section),
with the likelihood is maximized at $\alpha=1.00^{+2.38}_{-1.00}$
and $\beta=2.34^{+0.26}_{-0.24}$, from which one could
clearly see the marginal consistency between our fits and recent
measurements of three stellar VDFs (especially the SDSS DR5 VDF of
early-type galaxies).

What would be an appropriate interpretation of this disagreement
between the local stellar VDF and the lensing-based inferred VDF? In
order to answer this question we must quantitatively examine the
effects of the evolution in the velocity dispersion function.
Therefore, in the second model we adopt the number density and mass
evolution in the early-type galaxy population from the recent
studies of \citet{Faber2007,Brown2007}, which gave a decline in the
abundance by roughly a factor of two ($\nu_n=-1$) and a 20\%
increase in the velocity dispersion ($\nu_v=0.25$) for early-type
galaxies from $z=0$ to 1. Limits on the VDF shape parameters are
also shown in Table 1 and Fig.~2. For Sample A,
$\alpha=0.45^{+2.38}_{-0.45}$ and $\beta=2.55^{+0.28}_{-0.20}$ are
obtained at 68.3\% confidence level, while the constraints on the
individual VDF parameters are $\alpha=0.70^{+2.63}_{-0.70}$ and
$\beta=2.55^{+0.30}_{-0.24}$ for Sample B. The main features of
Fig.~2 may be summarized as follows. Firstly, comparing constraints
based on no-evolution and redshift-evolution models, one may clearly
see that the VDF parameters obtained from the redshift-evolution
model disagree with the respective value derived from no-evolution
model at $1\sigma$. More specifically, fits on the high-velocity
cut-off index reveal the better consistency between the solely
lensing-based VDF (assuming passive evolution) and the measured SDSS
DR5 VDF of early-type galaxies in the local universe. Hence the
evolution of the velocity dispersion function still significantly
affects lensing statistics if the lensing galaxies are of early type
\citep{Mitchell2005}. Secondly, both of the SDSS DR5 and SDSS DR12
measured stellar VDFs agree very well with the lens redshift
distribution, while the simulated local stellar VDF is disfavored at
high confidence levels ($>3\sigma$). Thirdly, it is of great
importance to take into account the effects of sample
incompleteness, given the lens redshift range of $0.234<z_l<1.00$
for Sample A (with the median value of $z_l=0.268$) and
$0.06<z_l<0.72$ for Sample B (with the median value of $z_l=0.208$).
Especially, for the Sample B we find that the lensing-based values
of ($\alpha, \beta$) are nearly equal to the corresponding stellar
values for the adopted SDSS DR5 VDF in the redshift-evolution
scenario. This is an argument in favour of the efficiency of lens
redshift test as a probe of the velocity dispersion function of
early-type galaxies. Therefore, our results indicate that sample
selection plays an important role in determining the VDF shape of
early-type population from the lens redshift test. Such findings,
which highlight the importance of considering galaxy evolution and
sample selection to better investigate the global properties of
early-type galaxies, have been extensively discussed in many
previous works focusing on improved constraints on the cosmological
parameters through strong lensing statistics \citep{Cao12a,Cao12b}.

\begin{figure*}
\centering
\includegraphics[width=12cm]{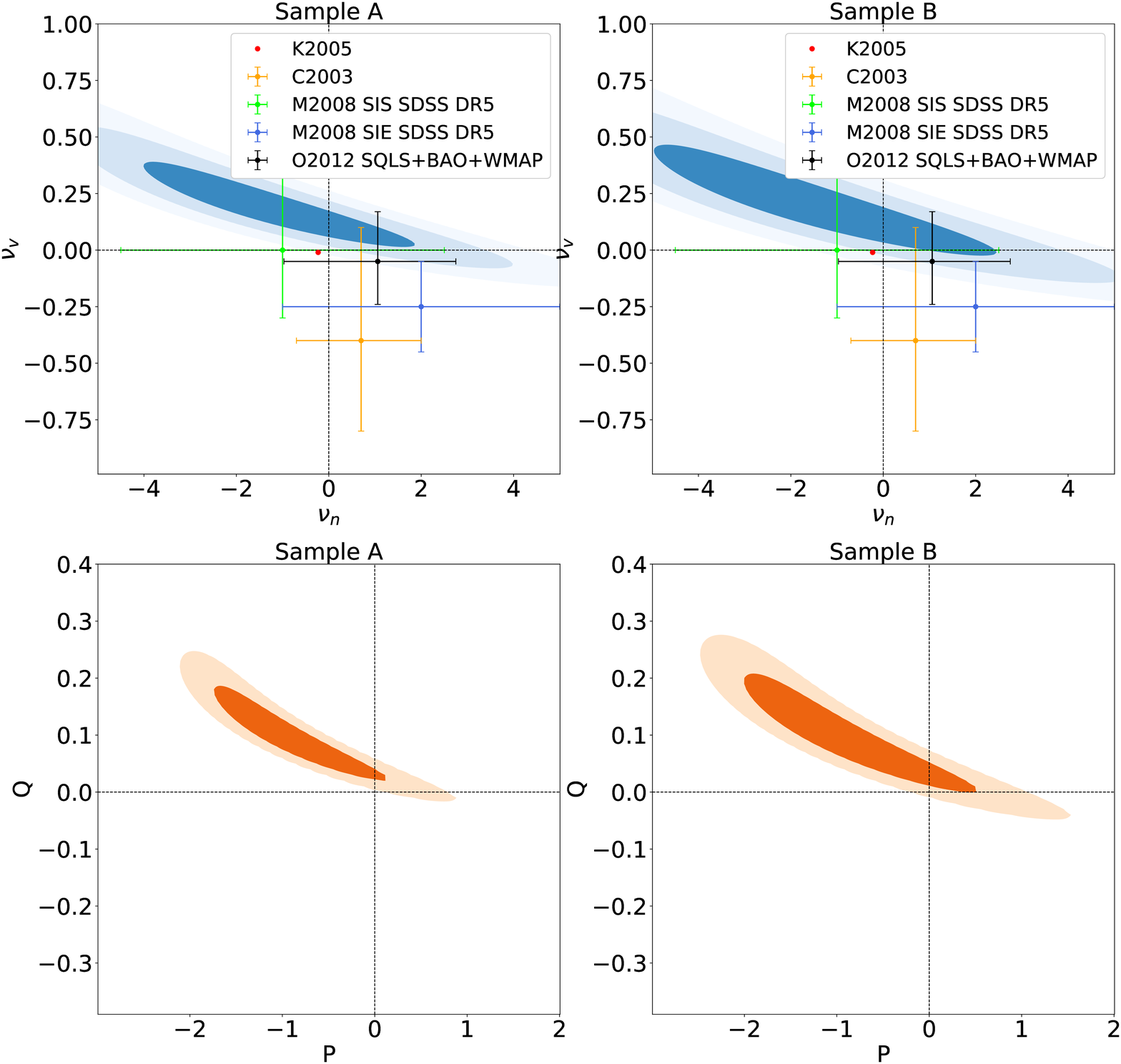}
\caption{Constraints on the redshift evolution of the VDF, i.e., the
number density evolution index $\nu_n$ ($P$) and the velocity
dispersion evolution index $\nu_v$ ($Q$) with two different lens
samples (Sample A and Sample B). The no-evolution model (black
dashed lines) and the evolution parameters obtained in the
literature are also added for comparison (see the text for more
details).} \label{fig:5}
\end{figure*}

\begin{table}
    \centering
    \caption{Summary of the constraints on the evolution parameters,
    based on the lens redshift distribution of the current strong
    lensing observations.}
    \label{tab:2}
    \begin{tabular}{l|l|l|llrrrr}
        \hline \hline
        VDF parameters $(\sigma_*, \alpha, \beta)$ & Data  & $\nu_n$  & $\nu_v$ \\
        \hline
        (161 km/s, 2.32, 2.67)  & Sample A & $-1.18^{+3.04}_{-2.82}$ & $0.18^{+0.21}_{-0.16}$ \\
        $(161\pm5 \kms$,  & Sample A  & $-1.02^{+3.15}_{-2.74}$ & $0.17^{+0.19}_{-0.16}$ \\
        $2.32 \pm 0.10, 2.67 \pm 0.07)$ & & & \\
        (161 km/s, 2.32, 2.67)  & Sample B & $-1.56^{+4.01}_{-3.39}$ & $0.20^{+0.26}_{-0.22}$ \\
        \hline
        VDF parameters $(\sigma_*, \alpha, \beta)$  & Data  & $P$  & $Q$ \\
        \hline
        (161 km/s, 2.32, 2.67)  & Sample A & $-0.87^{+0.99}_{-0.86}$ & $0.09^{+0.09}_{-0.07}$ \\
        (161 km/s, 2.32, 2.67)  & Sample B & $-0.88^{+1.39}_{-1.11}$ & $0.09^{+0.11}_{-0.09}$ \\
        \hline \hline
    \end{tabular}
\end{table}

\subsection{Constraints on the evolution of early-type galaxies}

Considering the redshift range of the lensing galaxies referring to
the distant universe (from $z=0.06$ to $z=1.0$), it is quite
necessary to stress another question, that is: \emph{Is it possible
to achieve a stringent constraint on galaxy evolution by using
gravitational lenses since redshift $z\sim 1.0$?} From this point of
view, in the framework of the VDF for local early-type galaxies
based on SDSS Data Release 5 \citep{Choi2007}, we assume that the
(characteristic) number density and velocity dispersion of lensing
galaxies evolves as a function of redshift according to Eq.~(6).

Notice that the previous studies always considered only the
evolution of the number density and the characteristic velocity
dispersion, assuming a constant shape of the inferred velocity
dispersion function \citep{Chae2003,Ofek2003}. In this analysis, the
measured value of the two evolution parameters are
$\nu_n=-1.18^{+3.04}_{-2.82}$ and $\nu_v=0.18^{+0.21}_{-0.16}$ when
the full lensing sample is taken into consideration. For Sample B,
the best-fit values and the 1$\sigma$ limits change to
$\nu_n=-1.56^{+4.01}_{-3.39}$, $\nu_v=0.20^{+0.26}_{-0.22}$. Fig.~3
shows the constraints in the $\nu_n-\nu_v$ plane. On the one hand,
our analysis reveals the strong degeneracy between the evolutions of
the velocity dispersion and the number density, which is supported by the
most degenerate direction in the evolution parameters
\citep{Oguri2012}. On the other hand, the inferred evolutionary
trends of the early-type VDFs are particularly interesting: lensing
statistics demand that for early-type population at redshift $z=1$,
there should be a decline in the number density by a factor of two
($\nu_n\sim-1$) compared with the present-day value. This implies
that dynamically, the population of lensing galaxies can be much
different from the present-day population. Meanwhile, such change in
number also requires a 20\% increase in the characteristic velocity
dispersion ($\nu_v=0.25$), following the general scenario of
high-redshift formation and passive evolution of early-type galaxies
reported by the recent studies \citep{Brown2007,Faber2007}. We
remark here that in the $\Lambda$CDM hierarchical structure
formation picture, the dark halo mass function (DMF) and the
velocity dispersion function (VDF) of galaxies evolves in cosmic
time as a consequence of hierarchical merging
\cite{White1978,Lacey1993}. In our analysis, the lensing-based VDF
evolution is strikingly similar to the prediction of the CDM
hierarchical structure formation paradigm, concerning the DMF from
N-body simulations and the stellar mass function (SMF) predicted by
recent semi-analytic models of galaxy formation. Interestingly, our
results are particularly in conflict with the results of several
galaxy surveys \cite{Fontana2006,Pozzetti2007,Marchesini2009}, which
support the stellar mass-downsizing evolution of galaxies
(apparently anti-hierarchical). The disagreement between the lensing
constraints and galaxy survey results is more apparent when the
large size difference between the samples is taken into
consideration.

One should bear in mind that the above results involve several
uncertainties and assumptions that act as systematic errors in our
analysis. Specifically, we consider the following sources of
systematic errors in a similar way as done in
\citet{Matsumoto2008,Cao12a}. Firstly, one possible source
of uncertainties inherent to our analysis are the uncertainties of
VDF parameters given by Eq.~(11). Therefore, based on the full lens
sample, we vary the characteristic velocity dispersion by
$\Delta\sigma_*=5$km/s and obtain the constraints on the shape of
early-type VDF in Table 1. Furthermore, while adopting the best-fit
VDF measurement in the SDSS DR5 as our fiducial model, we perform a
sensitivity analysis and investigate how galaxy evolution is altered
by introducing the uncertainties of $\alpha$, $\beta$ and
$\sigma_*$. The final results show that although the constraints
become relatively weak due to the uncertain stellar VDF measurements
in the local universe, central fits are almost the same. Secondly,
the parametrization of galaxy evolution is another important source
of systematic error on the final results. Although this problem has
been recognized long time ago, the most straightforward solution to
this issue is focusing on other well-known evolution models, which
have been widely used for analysis of statistical lensing in the
previous works \citep{Oguri2012}. Consequently, in the following
analysis we also perform fits on the second evolution model with a
constant shape of the VDF. Specially, the velocity
dispersion (as well as the number density) vary as a function of
redshift [Eq.~(7)] \citep{Ofek2003, Chae2010}. The simultaneous
constraints on the redshift evolution of the velocity dispersion
function $(P, Q)$ are shown in Table 2 and Fig.~3, with the best-fit
value of $P=-0.87^{+0.99}_{-0.86}$, $U=0.09^{+0.09}_{-0.07}$ (Sample
A) and $P=-0.88^{+1.39}_{-1.11}$, $Q=0.09^{+0.11}_{-0.09}$ (Sample
B). Again, our studies of galaxy evolution (based on the lens
redshift distribution of Sample A and Sample B) still prefer a
significant evolution in the number and mass of early-type galaxies
at $z\sim 1$.

\section{Discussions}

In this paper, we have used the strong lensing statistics (i.e., the
lens redshift distribution) of a well-defined sample of early-type
gravitational lenses extracted from a large collection of 157
systems to constrain the velocity dispersion function and the
evolution of early-type galaxies. By adopting a power-law model for
galactic potentials and the cosmological parameters determined by
the recent \textit{Planck} observations, we employ the lens redshift
test proposed in \citet{Kochanek1992,Ofek2003} to constrain the
velocity dispersion function (VDF) of early-type galaxies and its
evolution in a more complicated model. Our results have shown that
the population of early-type galaxies can be much different from the
present-day population (Sec. 3.1) and the lens redshift distribution
is a sensitive probe of galaxy evolution (Sec. 3.2).

One important issue is the comparison of our results with the
estimation of parameters in the power-law evolution model obtained
in the previous studies
\citep{Chae2003,Kang2005,Matsumoto2008,Oguri2012}. This is
illustrated in Fig.~3, which directly shows the evolution parameters
obtained in this analysis and in the literature. The red dot circle
denotes the best-fitted evolution parameters predicted by the
semi-analytic model in \citet{Kang2005}, while the orange cross
represents the 1$\sigma$ limits on the two evolutionary indices
derived by the observed image separations of 13 lenses from CLASS
\citep{Chae2003}. The green and blue crosses show the results
(1$\sigma$ uncertainties) with the WMAP's best-fitted $\Lambda$CDM
cosmology \citep{Matsumoto2008}, concerning the lens-redshift test
of the well-defined SDSS lens sample characterized by different
galaxy mass profiles (SIS and SIE lens for SDSS DR5). The black
cross denotes the recent measurements of the redshift evolution of
VDF based on the statistical analysis of the final sample from SQLS,
combined with external cosmological probes such as BAO and WMAP
\citep{Oguri2012}. We obtain the constraints on the evolution of the
characteristic number density and velocity dispersion which are in
broad agreement with these previous studies (especially the SIS
lenses for SDSS DR5) \citep{Matsumoto2008}. We also compare our
results with recent fits obtained in recent measurements of the
stellar mass function, based on the examination of galaxy
populations at $z\sim 1$. More specifically, although the effects of
the redshift evolution form has yet to be fully clarified, it was
found in \citet{Ilbert10,Matsuoka10,Brammer11} that the number
density of galaxies for a given stellar mass range can evolve by a
factor of two from $z=0$ to 1. Further papers have also noticed such
redshift evolution in the direct evolution measurement of the
velocity function (up to $z\sim1$), focusing on a scaling relation
between velocity dispersion, stellar mass, and galaxy structural
properties \citep{Bezanson11}. The above findings are in fact
compatible with our results based on the strong lensing statistics
(i.e., the redshift distribution of galactic-scale lenses).

The importance of galaxy evolution in strong lensing statistics was
also widely recognized \citep{Chae2003,Chae2010,Oguri2012}. Such
evolution scenario of early-type galaxies, which coincides with the
recent studies of the fundamental plane of lensing galaxies
\citep{Kochanek2000,Rusin2003}, is different from that obtained in
\citet{Chae2003}, with the best-fit value of $\nu_n$ being positive
and $\nu_v$ being negative in the hierarchical structure formation
theory. Our different conclusion might be due to the fact that our
lens sample is extended to a wide coverage of velocity dispersions
(98km/s$\leq \sigma \leq$396 km/s). As was found in the recent
analysis of the early-type VDFs from $z=1$ to $z=0$
\citep{Matsumoto2008,Chae2010}, the differential number density of
early-type galaxies would experience greater evolution at a higher
velocity dispersion. Specially, an increasingly large factor
($\geq$3) was reported for the galaxies at the largest velocity
dispersion end ($\sigma\geq$ 300 km/s) since $z=1$ \citep{Chae2010}.
However, the lens sample used by the previous lensing statistics is
restricted to early-type galaxies with typical velocity dispersions
and lower ($\sigma\leq 230-250$km/s), for which there is no
statistically significant change in number density from $z=0$ to
$z=1$ \citep{Oguri2012}. This explains why the no-evolution of the
early-type population, which has usually been used to constrain
cosmological parameters and test the properties of dark energy,
appears to be supported by different strong lensing statistics
\citep{Chae05,Mitchell2005}. On the other hand, the lens sample used
in this analysis is more complete in the source and lens redshifts,
with a better understanding of the selection function than that
adopted in \citet{Chae2003,Chae05}.

\begin{figure}
\centering
\includegraphics[width=7cm]{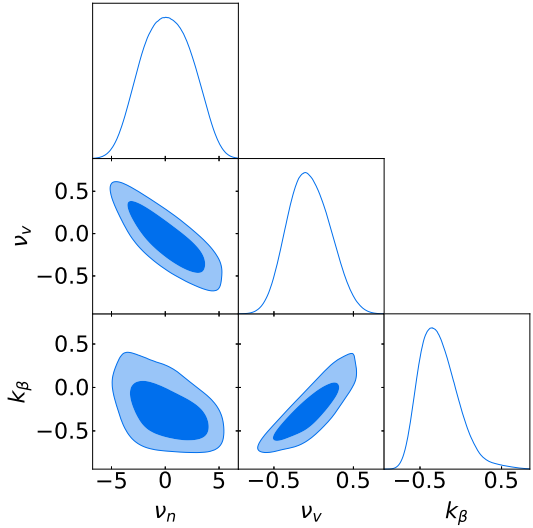}
\caption{Constraints on the redshift evolution of the VDF in a more
complicated evolution model with the full lens sample (Sample A).
The additional parameter $k_\beta$ is included to quantify the
evolution of the VDF shape.}
\end{figure}

Most studies of strong lensing statistics have used the galaxy
population with constant VDF shape for the number density of lenses
\citep{Chae2003,Ofek2003,Chae05,Matsumoto2008,Cao12a}. Following the
recent results indicating the differential redshift evolution in the
number density of galaxies with different velocity dispersions
\citep{Bezanson12,Montero17}, we also include an additional
parameter $k_\beta$ to describe the redshift evolution of the shape
of the velocity function \citep{Chae2010}
\begin{equation}
\begin{aligned}
\alpha\rightarrow \alpha\left(1+k_\beta\frac{z_l}{1+z_l}\right), \
\beta\rightarrow \beta\left(1+k_\beta\frac{z_l}{1+z_l}\right).
\end{aligned}
\end{equation}
$k_\beta>-1$ is required to guarantee the positivity of the VDF
shape parameters ($\alpha$, $\beta$). Note that such additional
parametrization, which is well consistent with the
intermediate-redshift VDFs steeper both at the low-velocity and
high-velocity ends \citep{Oguri2012}, could also effectively
describe the redshift dependence of the halo-mass function
suggesting stronger redshift evolution for larger velocity
dispersions \citep{Mitchell2005,Matsumoto2008}. The
constrains on individual evolution parameters are presented in
Fig.~4. It should be stressed that in the framework of a more
complicated model, the evolution of the VDF shape and the 1$\sigma$
limits is $k_\beta=-0.25^{+0.17}_{-0.29}$ for the full lensing
sample, which is marginally consistent with the fiducial
no-redshift-evolution case ($k_\beta=0$). However, the correlation
between $k_\beta$ and ($\nu_n, \nu_v$) is apparently indicated in
our analysis, i.e., a significant evolution of the VDF shape will
lead to a smaller value for the (characteristic) number density
evolution and a larger value for the velocity dispersion evolution
of lensing galaxies.

As a final remark, we point out that the lensing constraints on the
velocity dispersion function and the evolution of early-type
galaxies are quite competitive, compared with those from other
methods such as redshift survey of galaxies. However, our analysis
still potentially suffers from the limited size of the lens sample.
Meanwhile, strong lensing systems with high-redshift quasars ($z\sim
5$) acting as the background source will dramatically contribute to
investigating the effects of galaxy evolution on the lens
statistics, as was found in the recent works of \cite{Chae2003}.
With planned upgrades in next generation of wide and deep
sky surveys (such as the Large Synoptic Survey Telescope (LSST)), it
is possible to discover $10^5$ galactic-scale lenses in the near
future, with the corresponding source redshift reaching to $z\sim 6$
\citep{Collett2015}.

For this purpose we create mock data including $10^4$ strong lensing
systems on the base of realistic population models of elliptical
galaxies acting as lenses. Construction of the mock catalog
proceeded along the following steps \citep{Cao19,Cao20}: I) For the
purpose of calculating the sampling distribution (number density) of
lenses, we use the VDF of elliptical galaxies in the local Universe
derived from SDSS Data Release 5 \citep{Choi2007}. Meanwhile, in our
simulation we assume that neither the shape nor the normalization of
this function vary with redshift. The lens redshift of our simulated
sample, whose distribution is well approximated by a Gaussian with
mean $z_l=0.45$ and well consistent with the properties of the SL2S
sample, could reach to $z_l\sim 2$ \citep{Liu20a,Liu20b}. II) The
lens mass distribution is approximated by the singular isothermal
ellipsoids, while the simulated population of lenses is dominated by
galaxies with velocity dispersion of $\sigma_0=210\pm50$km/s. III)
The three angular diameter distances of the lensing systems (from
observer to lens, from observer to source, and from lens to source)
are calculated in the framework of a fiducial cosmological model
($\Lambda$CDM) from recent \textit{Planck} observations
\citep{Ade2016}. The effectiveness of our method could be seen from
the discussion of this question: \emph{Is it possible to achieve a
stringent measurement of the evolution of early-type galaxies?} In
the framework of a generalized evolution model given the evolution
of VDF shape, one can expect the evolutions of the number density
and the characteristic velocity dispersion to be estimated with the
precision of $\Delta \nu_n=0.085$ and $\Delta \nu_v=0.006$. The
resulting constraint on the redshift evolution of the VDF shape
becomes $\Delta k_\beta=0.012$, when the differential redshift
evolution in the number density of galaxies with different velocity
dispersions is taken into consideration. Now the final question is
\emph{Is it possible to confirm or falsify alternative semi-analytic
models of galaxy formation?} In particular, many galaxy surveys
suggest stellar mass-downsizing (apparently anti-hierarchical)
evolution of galaxies
\citep{Fontana2006,Pozzetti2007,Marchesini2009}, although there are
results that do not particularly support such evolution scenario
\citep{Brown2007,Faber2007}. The most striking conclusion of these
works is the emergence of the considerable evolutions of the number
density and the characteristic velocity dispersion, i.e., $(\nu_n,
\nu_v)=(-1, 0.25)$ for hierarchical model. Given the fact that the
precision is inversely proportional to the $\sqrt{N}$ (where $N$ is
the number of systems), with $10^4$ strong lensing systems one can
effectively differentiate between the hierarchical and
anti-hierarchical models at very high confidence $>5\sigma$.
Summarizing, when such a lens sample -- which increases the current
lens size by orders of magnitude -- is available, one could expect
the most stringent lensing constraints on the formation and
evolution of early-type galaxies, from the redshift distribution of
gravitational lenses.

\section*{Acknowledgments}

This work was supported by National Key R\&D Program of China No.
2017YFA0402600; the National Natural Science Foundation of China
under Grants Nos. 12021003, 11690023, 11633001, and 11920101003;
Beijing Talents Fund of Organization Department of Beijing Municipal
Committee of the CPC; the Strategic Priority Research Program of the
Chinese Academy of Sciences, Grant No. XDB23000000; the
Interdiscipline Research Funds of Beijing Normal University; and the
Opening Project of Key Laboratory of Computational Astrophysics,
National Astronomical Observatories, Chinese Academy of Sciences.
This work was initiated at Aspen Center for Physics, which is
supported by National Science Foundation grant PHY-1607611. This
work was partially supported by a grant from the Simons Foundation.
M. Biesiada is grateful for this support.

%%%%%%%%%%%%%%%%%%%%%%%%%%%%%%%%%%%%%%%%%%%%%%%%%%
\section*{Data Availability Statements}

The data underlying this article will be shared on reasonable
request to the corresponding author.


%%%%%%%%%%%%%%%%%%%%%%%%%%%%%%%%%%%%%%%%%%%%%%%%%%

% Don't change these lines
\bsp    % typesetting comment
\label{lastpage}
\end{document}